\documentclass[sigconf,authorversion,nonacm]{acmart}

\usepackage{subcaption}
\usepackage[capitalise]{cleveref}
\usepackage{tabularx}

\newcolumntype{L}{>{\raggedright\arraybackslash}X}

\begin{document}

\title{A Comprehensive Benchmarking Analysis of Fault Recovery in Stream Processing Frameworks}

\author{Adriano Vogel}
\orcid{0000-0003-3299-2641}
\affiliation{%
    \institution{JKU/Dynatrace Co-Innovation Lab\\Johannes Kepler University Linz}%
    \city{Linz}%
    \country{Austria}%
}
\email{adriano.vogel@jku.at}

\author{Sören Henning}
\orcid{0000-0001-6912-2549}
\affiliation{%
  \institution{JKU/Dynatrace Co-Innovation Lab\\Johannes Kepler University Linz}%
  \city{Linz}%
  \country{Austria}%
}
\email{soeren.henning@jku.at}

\author{Esteban Perez-Wohlfeil}
\orcid{0000-0002-3572-686X}
\affiliation{%
    \institution{Dynatrace Research\\Dynatrace LLC}%
    \city{Linz}%
    \country{Austria}%
}
\email{esteban.wohlfeil@dynatrace.com}

\author{Otmar Ertl}
\orcid{0000-0001-7322-6332}
\affiliation{%
    \institution{Dynatrace Research\\Dynatrace LLC}%
    \city{Linz}%
    \country{Austria}%
}
\email{otmar.ertl@dynatrace.com}

\author{Rick Rabiser}
\orcid{0000-0003-3862-1112}
\affiliation{%
    \institution{LIT CPS Lab\\Johannes Kepler University Linz}%
    \city{Linz}%
    \country{Austria}%
}
\email{rick.rabiser@jku.at}

\begin{abstract}
Nowadays, several software systems rely on stream processing architectures to deliver scalable performance and handle large volumes of data in near real-time. Stream processing frameworks facilitate scalable computing by distributing the application's execution across multiple machines. Despite performance being extensively studied, the measurement of fault tolerance—a key feature offered by stream processing frameworks—has still not been measured properly with updated and comprehensive testbeds.
Moreover, the impact that fault recovery can have on performance is mostly ignored. This paper provides a comprehensive analysis of fault recovery performance, stability, and recovery time in a cloud-native environment with modern open-source frameworks, namely Flink, Kafka Streams, and Spark Structured Streaming. Our benchmarking analysis is inspired by chaos engineering to inject failures. Generally, our results indicate that much has changed compared to previous studies on fault recovery in distributed stream processing. In particular, the results indicate that Flink is the most stable and has one of the best fault recovery. Moreover, Kafka Streams shows performance instabilities after failures, which is due to its current rebalancing strategy that can be suboptimal in terms of load balancing. Spark Structured Streaming shows suitable fault recovery performance and stability, but with higher event latency. Our study intends to  (i) help industry practitioners in choosing the most suitable stream processing framework for efficient and reliable executions of data-intensive applications; (ii) support researchers in applying and extending our research method as well as our benchmark; (iii) identify, prevent, and assist in solving potential issues in production deployments.  

\end{abstract}


\maketitle

\section{Introduction}

\textit{Stream processing} refers to a computing paradigm and architecture pattern for building distributed, event-based software systems that process massive amounts of data in near real time~\cite{Hirzel2013}. 
Today's stream processing paradigms enable software engineers to model their applications' business logic as high-level representations in directed acyclic graphs (DAGs), without explicitly defining the physical execution plan. This high level of abstraction is provided by industry-grade, open-source stream processing frameworks such as Spark~\cite{Armbrust2018}, Flink~\cite{Carbone2015}, or Kafka Streams~\cite{Wang2021}.
Such frameworks support software engineers in building highly scalable and efficient applications that process continuous data streams of massive volume.
They provide high-level APIs and domain-specific languages to define the processing logic as directed acyclic processing graphs that filter, transform, aggregate, and merge data streams.

Over the last decade, several works have been published that evaluate the performance of stream processing frameworks~\cite{SEAA2023,ICPE2024} or optimizations~\cite{Hirzel2013,Herodotou2020}.
Beyond measuring performance metrics such as throughput and latency, fault tolerance is a critical requirement for continuously operating production systems. Failures can occur unpredictably across various levels, from physical infrastructure to software layers. Stream processing systems, designed for continuous, low-latency processing, demand swift recovery mechanisms to tolerate and mitigate failures effectively. 

It is notable that a relevant number of studies reported fault tolerance as challenging for stream processing~\cite{SEAA2023}. Although fault tolerance guarantees such as at-least-once or exactly-once are already provided by modern stream processing frameworks~\cite{Fragkoulis2023} and partially evaluated in the literature~\cite{vanDongen2021a,Wang2022}, there are still many open questions for stream processing in scalable industry-grade deployments: i) With the new developments of fault tolerance features, how fast do the executions recover from failures? ii) What is the failures' impact on the application performance? iii) What is the impact of more critical failures such as correlated failures of many worker instances (entities executing the stream processing applications)? iv) How stable is the fault recovery across recurrent failures (we call it stability)?
        
Considering the open questions of stream processing frameworks' resilience to failures, we aim to extend the current methods for evaluating fault recovery and provide updated insights from conducting comprehensive experimental analysis. Particularly, this paper provides the following contributions to industry practitioners and the research community:

\begin{itemize}

    \item[--] An extension of ShuffleBench~\cite{ICPE2024} for fault tolerance measurements in the different state-of-the-art stream processing frameworks, tooling to automate benchmarks' execution, and data collection in Kubernetes-based cloud environments. 
    
    \item[--] A comprehensive methodology for measuring fault recovery. The methodology covers stream processing's most relevant metrics (throughput and latency)~\cite{SEAA2023} and new metrics to improve fault recovery measurements (fault recovery impact on performance, stability, and recovery time).

    \item[--] A method inspired by chaos engineering that enables the injection of different types of recurrent failures in stream processing experiments.
    
    \item[--] A method to automate fault recovery time measurements, available as supplementary material~\cite{ReplicationPackage}.
    
    \item[--] An experimental analysis covering application metrics, system resources utilization, and fault recovery. Our evaluation covers the open-source stream processing frameworks Flink, Kafka Streams, and Spark Structured Streaming due to their industry acceptance and academic relevance~\cite{SEAA2023}.
\end{itemize}

The experimental data and an extended replication package of ShuffleBench~\cite{ICPE2024} for fault tolerance measurements are available as supplementary material~\cite{ReplicationPackage}.

\section{Background}\label{sec:background}

This section briefly introduces the fundamental concepts of modern stream processing frameworks and fault tolerance.

\subsection{Distributed Stream Processing}\label{sec:background:stream-processing}

Stream processing frameworks perform operations such as filterings, transformations, or aggregations in near real time on continuous streams of data~\cite{Hirzel2013}.
State-of-the-art frameworks are designed for high throughput and low-latency processing, while also scaling with massive amounts of data~\cite{Fragkoulis2023,ICPE2024}.
To address these requirements, they run in a distributed fashion on commodity hardware, nowadays often in managed cloud environments.

A key advantage of stream processing frameworks is that they provide dataflow models that abstract aspects such as cluster management, state management, and time semantics from their users~\cite{Sax2018}. 
The frameworks allow the initiation of multiple worker instances across various compute nodes with multiple threads, each instance handling a distinct portion of the data.

While the isolated processing of data records remains unaffected by the assignment of data portions to worker instances, processing that depends on previous data records, such as aggregations, requires state management.
Similar to the MapReduce~\cite{Dean2008} programming model, keys are assigned to records before a stateful operation. This allows the stream processing frameworks to route all records with the same key to the same instance, where state synchronization among instances can be avoided.
When a processing operator modifies the key of a record and a subsequent operator performs a stateful operation, the framework divides the dataflow graph into subgraphs that can be independently processed by different worker instances. Popular stream processing frameworks include Apache Flink~\cite{Carbone2015}, Apache Kafka Streams~\cite{Wang2021}, and Apache Spark with its Structured Streaming engine~\cite{Armbrust2018}.

\subsection{Fault Tolerance}\label{sec:background-fault-tolerance}

Achieving consistency in stream processing has posed a longstanding research challenge, which is partly attributed to the lack of a formal problem specification~\cite{Wang2021}. Although executions can be subject to many failures, we expect computing systems to produce correct results despite the failures.
Such a requirement instigated research on fault tolerance, where notable advances achieved in distributed systems are applied to stream processing~\cite{Wang2021}.

In the absence of fault tolerance, stream processing applications would be forced to restart data processing from scratch whenever there is a loss due to a failure. Processing semantics guarantees describe how a system is affected by failures. Various guarantees are provided to accommodate the varying fault tolerance needs of different use cases~\cite{Fragkoulis2023}. Relevant terms are at-most-once semantics (AMOS), at-least-once semantics (ALOS), and exactly-once semantics (EOS). 
Given that the majority of stream processing applications require reliable executions with strong semantics guarantees, we have shifted our focus from AMOS, which can be considered obsolete, to ALOS and EOS~\cite{Fragkoulis2023}.

\paragraph{At-least-once semantics} ALOS ensures that the system maintains consistent results even in the event of failures, preventing data loss. However, during recovery, duplicate output may occur as records might be processed more than once~\cite{Fragkoulis2023}. The frequency of such duplicates largely depends on the system's implementation, the type of failure encountered, and the timing of the failure during execution. Additionally, handling potential duplicates can be accomplished externally, for instance, through downstream systems filtering duplicates or database overwrites. With effective implementations, the fault tolerance guarantee provided by ALOS can be comparable to EOS.

\paragraph{Exactly-once semantics} EOS refers to the ability of the system to process the records exactly once without lost or duplicated results even under failures. Therefore, EOS requires additional implementation and has a performance impact on records processing~\cite{vanDongen2021a}.

In this paper, we focus on the semantic guarantees of the system output, as these are most relevant for the majority of use cases and also encompass the guarantees for internal state updates. 
The stream processing frameworks provide different recovery semantic guarantees. This is important because the required guarantee varies according to the stream processing application and its use cases. 

In some scenarios, guaranteeing exactly-once delivery is of paramount importance, such as avoiding duplicate alerts in an observability platform. In other cases, at-least-once with rarely duplicated outputs is not equally detrimental, e.g., an information dashboard that receives twice the same CPU metric of a timestamp. We intend to measure the impacts of the stream processing frameworks when running with the fault recovery guarantees. Importantly, correct fault recovery is a crucial component of fault tolerance. While fault recovery focuses on restoring normal operation after a failure, fault tolerance aims to prevent downtime for high availability.

\section{Research Landscape}
This section overviews the related literature and we discuss existing research and analysis gaps.

\subsection{State-of-the-art}\label{sec:background:benchmarking}

Recent publications provide experimental evaluations of fault recovery in stream processing~\cite{SEAA2023}. Although many studies consider failures of worker instances~\cite{Lu2014,Qian2016,Lopez2016}, the most updated and relevant is~\cite{vanDongen2021a}. It dives into the implementation, performance, and efficiency of fault recovery in four stream processing frameworks: Spark Streaming, Flink, Spark Structured Streaming, and Kafka Streams. In addition, ~\citet{vanDongen2021a} evaluate the behavior of these frameworks under different types of faults and settings, including master failure with and without high-availability setups, driver failures for Spark frameworks, worker failure with or without exactly-once semantics, and application failures.
 
\citet{Wang2022} propose a taxonomy of fault tolerance in stream processing. They also propose an evaluation framework tailored for fault tolerance, demonstrating experimental results on two representative open-source stream processing frameworks. Potential limitations in the current approaches are also discussed.

\citet{theodorakis2021} discuss the challenges and solutions in making single-node stream processing fault-tolerant while maintaining high performance. Due to the limited I/O bandwidth of a single node, it becomes infeasible to persist all stream data and operator state during execution. To address this issue, \textit{Scabbard} was proposed as the first single-node framework that supports exactly-once fault tolerance. However, the scalability of Scabbard can still be somewhat limited to a single node.

\citet{Wang2021} demonstrate how Kafka Streams works on large-scale deployments and performance insights exhibiting its flexible and low-overhead trade-offs. A recent paper from~\citet{siachamis2024} focuses on complementary aspects of fault tolerance, exploring three approaches for checkpoint protocols (coordinated, uncoordinated, and communication-induced) aimed at fault tolerance in stream processing. The analysis delves into the approaches' theoretical strengths and weaknesses. Their findings validate the preference for the coordinated approach and illustrated instances where the uncoordinated approach provides performance gains.

Considering that many optimizations can be done for fault tolerance~\cite{Hirzel2013}, \citet{su2021} address the issue of recovering from correlated failures in large-scale clusters. The authors propose an incremental and query-centric recovery paradigm. They formulate the problem of recovery scheduling under correlated failures and design algorithms to optimize the recovery latency.

\subsection{Research Gaps}\label{sec:gaps}

While fault recovery in stream processing has been addressed in various related studies, our understanding is that the methods and insights provided in the literature offer limited coverage. We believe that the following are relevant research gaps in the state-of-the-art regarding fault recovery on distributed stream processing: 

\begin{enumerate}
    \item Considering the continuous development of fault tolerance features and optimizations by the frameworks' open-source communities~\cite{Fragkoulis2023}, there is a need for updated measurements on frameworks' executions facing worker instances' failures.

    \item Few studies compare how different frameworks recover worker instances and those that do focus on failures affecting a single instance~\cite{vanDongen2021a,Wang2022}. However, a large physical machine crash impacts several virtual machines or pods. What can happen is that more than one worker instance fails. However, the impact of such a more critical failure on running stream processing applications remains unclear. 

    \item The existing literature only covers scenarios where failures occur a single time~\cite{vanDongen2021a,Wang2022}. However, in real-world industry deployments, failures happen recurrently and unpredictably. One can assume that the frameworks' recovery will always be the same, but we observed that such an assumption could have weak coverage with the reality of large deployments. This occurs mostly because the impact that a given failure has in the applications varies according to the exact moment that the failure happens (e.g., how far from the last saved state checkpoint), what working instances are affected by the failure, and the effectiveness of the subsequent recovery~\cite{geldenhuys2022}. Therefore, the level of fault recovery stability across recurrent failures demands more research.    
\end{enumerate}

\section{Experimental Setup}\label{sec:setup}

\Cref{{sec:gaps}} described existing research gaps. Here we show our experimental setup to extend the fault recovery measurements. 

\subsection{Benchmark}\label{sec:benchmark}

\emph{ShuffleBench}~\cite{ICPE2024} is a benchmark for evaluating the performance of modern stream processing frameworks. It focuses on use cases where stream processing frameworks are mainly employed for \emph{shuffling} (i.e., re-distributing) data records to perform state-local aggregations, while the actual aggregation logic is considered as a black-box software component.
ShuffleBench is inspired by requirements for near real-time analytics of a large cloud observability platform, but it is highly configurable to allow domain-independent evaluations. It comes as a ready-to-use open-source software\footnote{\url{https://github.com/dynatrace-research/ShuffleBench}} utilizing existing Kubernetes tooling and providing implementations for different state-of-the-art stream processing frameworks.

\begin{figure}
    \centering
    \includegraphics[width=\linewidth]{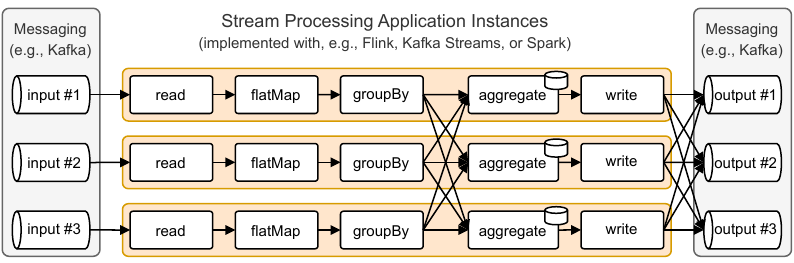}
    \caption{The ShuffleBench dataflow architecture of a stream processing application with three worker instances~\cite{ICPE2024}.}
    \label{fig:architecture}
\end{figure}

ShuffleBench adopts and extends the benchmarking framework Theodolite~\cite{EMSE2022} to automate the benchmark execution in Kubernetes-based cloud environments.
\Cref{fig:architecture} depicts our dataflow architecture for a corresponding stream processing application.
Data records are read from a messaging system (e.g., Kafka), assigned to keys (zero to many), and shuffled such that all records having the same key are forwarded to the same worker instance. There, a key-specific custom aggregation is performed, and an output event is created that is then written back to Kafka. We refer to our previous work for a detailed description of ShuffleBench~\cite{ICPE2024}.

\begin{figure}
    \centering
    \includegraphics[width=\linewidth]{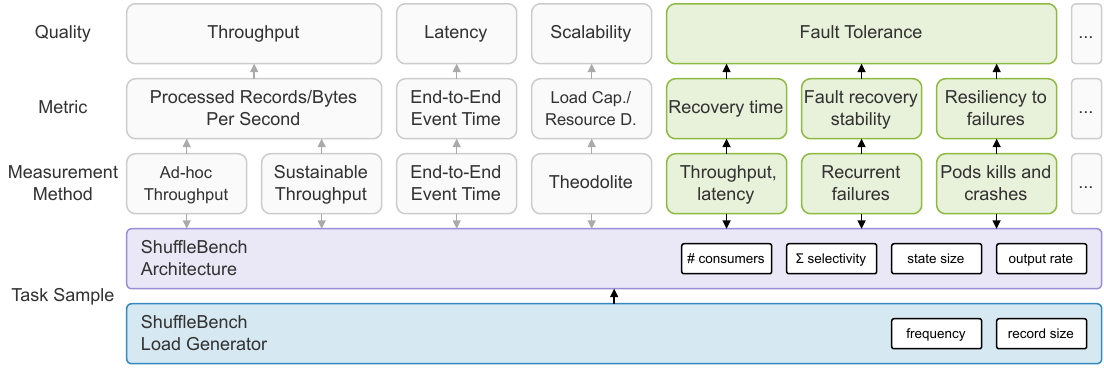}
    \caption{Overview of ShuffleBench benchmark components. Green components are extensions to our previous work~\cite{ICPE2024}.}
    \label{fig:components}
\end{figure}

ShuffleBench has a flexible architecture and supports several parametric configurations. In this work, we extend ShuffleBench for fault tolerance measurements.
\Cref{fig:components} provides an overview of \emph{ShuffleBench's} components and their extension for fault tolerance (see the scenarios and metrics in~\Cref{sec:method}).
We conduct our experimental evaluation in a Kubernetes cluster managed by the Elastic Kubernetes Service of Amazon Web Services with the following pods running in dedicated cloud instances: 2 load generators (\emph{m6i.xlarge} instances), 2 Kafka brokers (\emph{m6i.2xlarge}), a manager node used on Flink and Spark  (\emph{m6i.large}), and 8 worker instances pods with 1 CPU, 3 GB of RAM (running on \emph{c5.large} instances). 

In our experiments, we set up 20\,000 real-time consumers that all have the same selectivity, which sum up to 50\,\%, meaning that each record is forwarded to $0.5$ consumers on average. Each consumer emits an output event for every record received. To consume and output results, 40 input and output Kafka partitions were used to maintain a suitable parallelism level that minimizes overheads. We run each experiment for 56~minutes in such a way that the benchmarks run for long enough to extract meaningful quantitative data, test with recurrent failures injected, and have sufficient time for every subsequent recovery. 

\subsection{Frameworks Deployment}

We configured the frameworks on ShuffleBench deployment and parameters described in \cref{sec:benchmark}. To support fault tolerance, Flink (v. 1.17) and Spark Structured Streaming (v. 3.4) had checkpointing enabled for at-least-once semantics (ALOS), see more in \cref{{sec:background-fault-tolerance}}. Kafka Streams (v. 3.5) by default supports ALOS. 
Fault tolerance on Flink and Spark Structured Streaming involves periodic state backup via checkpointing, demanding a reliable filesystem to store state copies~\cite{vanDongen2021a}. We used in our setup an Amazon Elastic File System (EFS) volume mounted in the pods.

Kafka Streams provides fault recovery with its strong integration with Kafka. Kafka acts as a messaging layer to safeguard state information. In-memory state stores are associated with a replicated changelog topic within Kafka, which tracks state updates. Whenever a task interacts with the store, it maintains a changelog. If a task encounters failure, the system automatically restarts it on one of the remaining worker instances. The state stores can be restored by replaying their changelog, ensuring data consistency. Kafka Streams provides a configuration to enable exactly-once (\textit{exactly\_once\_v2}), which was enabled to evaluate exactly-once. Exactly-once in Flink can be enabled in a similar way by enabling the \textit{DeliveryGuarantee} configuration to \textit{exactly\_once}.

Flink's checkpointing interval and Kafka Streams' commit interval were set to 2 seconds. Although such value can be tuned~\cite{geldenhuys2022}, it is comparable to Spark executions' micro-batching interval after the warmup in our setup. This is relevant for having executions more comparable because Spark Structured Streaming saves a checkpoint when the batch is completed.
Flink was also configured to run in the reactive mode, which is a modern feature necessary for supporting all the failures injected in our experiments.

Write-ahead-logs (WAL) was enabled in Spark Structured Streaming. Moreover, the number of SQL shuffle partitions was set to 16 to balance between the parallelism level and the resources. The Spark manager pod needed additional containers, the spark driver (execution manager), and Kafka offset committer that we implemented to support application metrics measurements in a comparable way to the other frameworks.

\section{Experimental Method}\label{sec:method}

This section describes our proposed method to evaluate fault recovery on stream processing.

\subsection{Goals}
Considering the best practices and the existing research gaps described in \cref{sec:gaps}, the following are the goals of our research method: i)~Create comparable deployments across different frameworks with similar throughput processing capabilities; ii)~Measure with comprehensive evaluation how the frameworks react to failures, covering periodic and recurrent failures of different types; iii)~Have evaluations that are realistic to real-world scenarios, focusing on the most likely failures affecting worker instances.

\subsection{Failure Injection Scenarios} \label{sec:scenarios}

Our experimental scenarios were inspired by chaos engineering~\cite{Basiri2016} using Chaos Mesh\footnote{Chaos Mesh is an open-source platform: \url{https://github.com/chaos-mesh/chaos-mesh}.} to inject failures in two scenarios:

\paragraph{Pods kill:} Random pods executing worker instances are deleted. When the pods are no longer accessible, both the records they were processing and their internal state are lost. Therefore, the frameworks trigger recovery from such a failure. Deployed with Kubernetes, new pods will be created and added to the pool of worker instances within an interval of seconds (usually taking from 2 to 10 seconds). We experimented with killing 1, 2, and 4 worker pods at once. Such values were determined to simulate simple and critical failure scenarios. The failures are recurrently injected every 5 minutes to evaluate the fault recovery stability. This interval was defined for a reasonable recovery time between failures and also to avoid unnecessary long executions that would waste resources.
     
\paragraph{Pods crash:}  Random worker instances pods face an operating system crash. This is expected to further test the resilience because the availability of resources is impacted. Contrasting with pods kill, in the pods crash failure no new pods are created to replace the failed ones. This failure also is expected to cause loss in the internal state of the worker instances affected, demanding failure recovery. Every failure impacts two random worker pods, which is a value set to represent a critical failure. The failures are recurrently injected every 10 minutes to evaluate the fault recovery stability and the failure remains for 5 minutes (value set according to feedback of industry partners). Then, executions can resume the normal state. Moreover, 5 minutes is a suitable value for creating and adding new virtual machines to the pool~\cite{su2021}.

\subsection{Metrics} \label{sec:metrics}

We collected several metrics of the benchmarking application. The metric to measure processing speed is throughput, measured in records processed per time unit and collected every second as a moving average of the last 5 seconds. A higher throughput is usually desired in many use cases.  
We measured input throughput (the rate of records consumed from the Kafka \emph{input} topic, see~\cref{fig:architecture}) and output throughput (the rate of records written to the Kafka \emph{output} topic).
We also measure event latency (lower is better) as an average of a 10-second window. We show latency percentiles to measure the distribution of events times: the median (p50), 90\textsuperscript{th} percentile (p90), and 99\textsuperscript{th} percentile (p99). 
We also measure the consumer lag corresponding to the number of records buffered in the input topic.

The fault recovery on stream processing is measured with the metrics: Fault recovery performance, the impact of fault tolerance guarantees, stability, and fault recovery time (see \Cref{fig:components}). We also present the following system monitoring indicators collected every 2 seconds from the cloud instances that run the worker pods: CPU usage, memory utilization, and network traffic.\footnote{To enhance visual clarity, the CPU usage and network traffic of worker nodes are displayed using a 10-value moving window average.}

The method described here is applied in \cref{sec:characterization,sec:stability,sec:recovery-time} to conduct measurements and extract insights. In this paper, we show the most insightful results, complementary results can be found in the supplementary material~\cite{ReplicationPackage}.

\section{Fault Recovery Characterization}\label{sec:characterization}

This section demonstrates how the injected failures impact the frameworks' executions.
The results cover application metrics and resource utilization (see more in \cref{sec:metrics}). This section intends to demonstrate the potential impact of faults, their recovery on the different stream processing frameworks, and their Quality of Service (QoS).
For visual clarity, The figures and plots provided for visual inspection depict either (i) the first three failures injected in the case of pod kills and (ii) the first failure followed by the frameworks' response in the case of pod crashes. \Cref{sec:stability} extends the characterization with results from up to 10 injected failures. 

\subsection{ALOS Recovery Characterization}

\begin{figure*}[htbp]
  \centering
  \begin{subfigure}[htbp]{1\textwidth} 
    \centering
    \includegraphics[width=\textwidth]{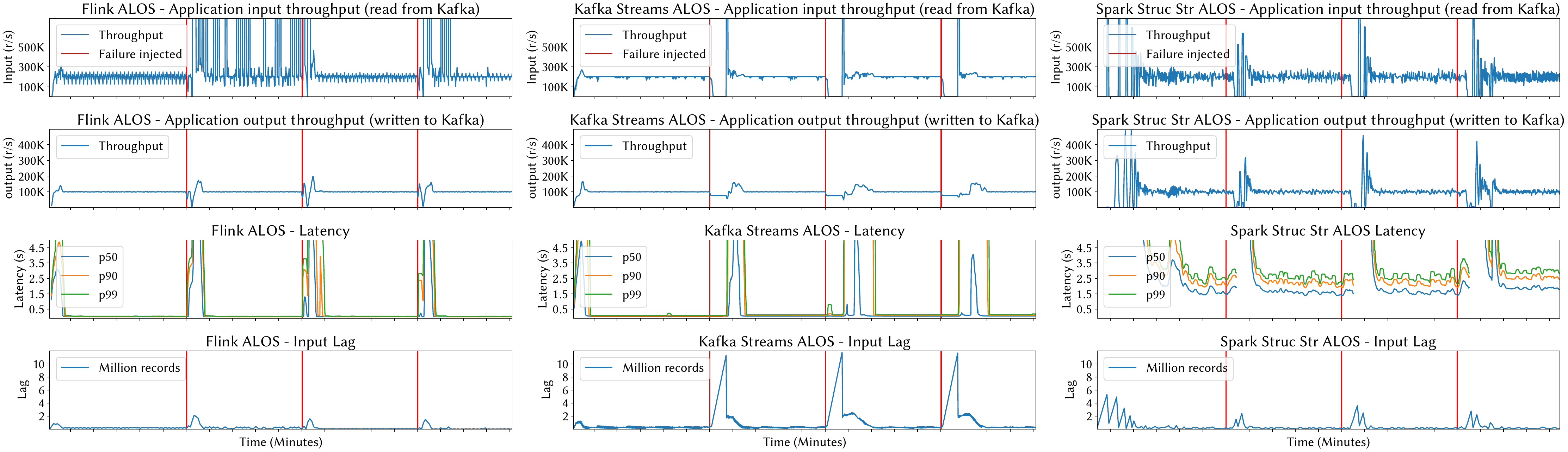}
    \caption{Fault recovery}
    \label{fig:characterization1podkill_performance}
  \end{subfigure}
    \begin{subfigure}[htbp]{1\textwidth} 
    \centering
    \includegraphics[width=\textwidth]{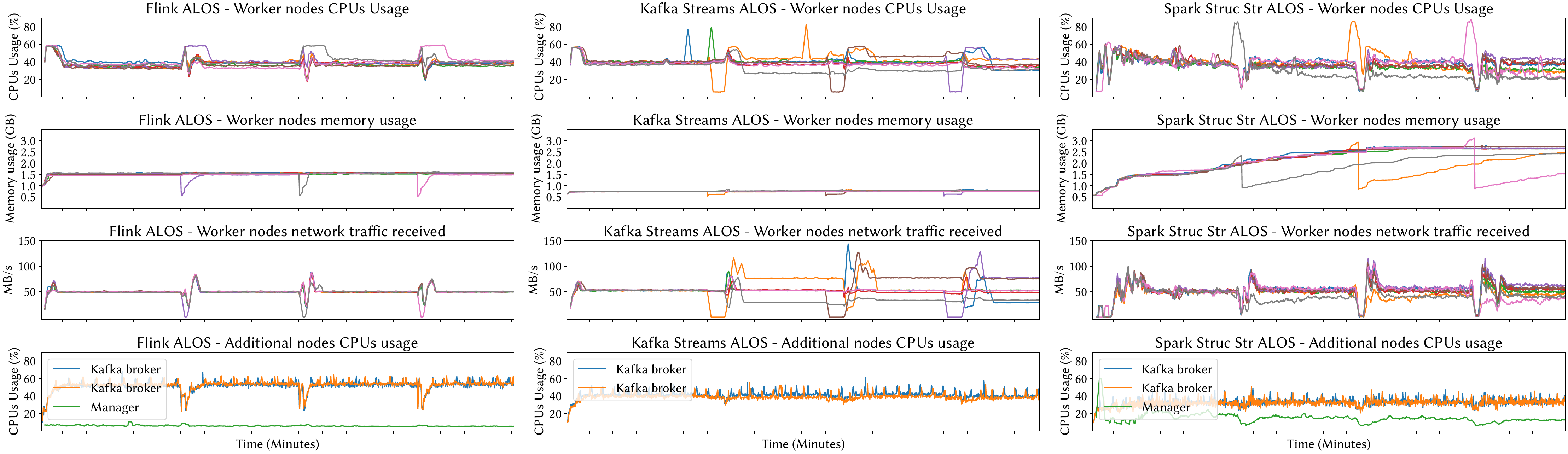}
    \caption{Resources utilization}
    \label{fig:characterization1podkill_resources}
  \end{subfigure}
  \caption{ALOS fault recovery characterization: Random pods executing worker instances are killed, repeated 3 times.}
  \label{fig:characterization1podkill}
\end{figure*}

\Cref{fig:characterization1podkill} shows executions with the different frameworks with pods executing worker instances being killed (represented by the red line). 
It is also important to note that the plots show different metrics (see y-axis labels) over time. When the failures occur, the worker instance affected is lost. Then, the Kubernetes creates a new pod, which is added to the pool of worker instances. Moreover, the stream processing frameworks start scheduling computations to the new worker instances. This is achieved in Flink and Kafka Streams by rebalancing the partitions among the worker instances. In Spark Structured Streaming, the new worker instances start receiving tasks to compute.  

\textbf{Fault~recovery~observation~1}: The \textit{frameworks recover correctly from the failures}. However, when comparing the frameworks, it is noticeable that Flink executions recover quicker from failures. This observation is evinced by a shorter downtime in the output throughput and a faster latency recovery. \textbf{Performance observation 1} is that after failures, \textit{Flink occasionally shows fluctuations in the input throughput}. To the best of our knowledge, this observation has not been previously reported. Such fluctuations in the input throughput after failures can be due to the variation of the failures instant relative to the last saved checkpoints. Considering that these variations do not significantly impact the output throughput or the latency, further analysis is left for future work.

\textbf{Performance observation 2}: \textit{Kafka Streams achieves a stable throughput with and without failures of one worker instance pod, but Kafka Streams' executions are subject to unstable latencies}. 
The Kafka Streams' partition assignment strategy triggers a rebalance after failures, causing its executions to accumulate more lag. Rebalancing significantly increases event latency and shows fluctuations in the worker machines' CPU usage and network traffic, indicating a potential imbalance. In fact, Kafka Streams' performance following failures is mostly impacted by its partition assignment strategy, which unevenly distributes partition to the consumer group's worker instances. Some worker instances may be assigned to more partitions or partitions with more accumulated lag. Then, some workers become overloaded while others are underutilized.\footnote{Further analysis and potential optimizations in our follow-up paper~\cite{vogel2024}.}

\textbf{Performance observation 3}: \textit{Spark Structured Streaming's dynamic micro-batch intervals stabilize after a few minutes on suitable configurations}. Dynamic micro-batching is noticeable in the first few minutes of execution due to fluctuations in throughput and lag, caused by an overly large micro-batch interval that results in more data being consumed and less frequent batches being triggered. 
Spark's self-regulated micro-batch interval stabilizes at around 2 seconds per batch after a few minutes of execution. This configuration sustains the input throughput while maintaining reasonable latency. In our experience, the measured p50 latency is comparable to the micro-batch interval. Still, Spark's latency is significantly higher than the other frameworks that process the records based on events instead of micro-batches. This insight is consistent with the related literature~\cite{vanDongen2021a}.

\begin{figure*}[htbp]
  \centering
  \begin{subfigure}[htbp]{1\textwidth} 
    \centering
    \includegraphics[width=\textwidth]{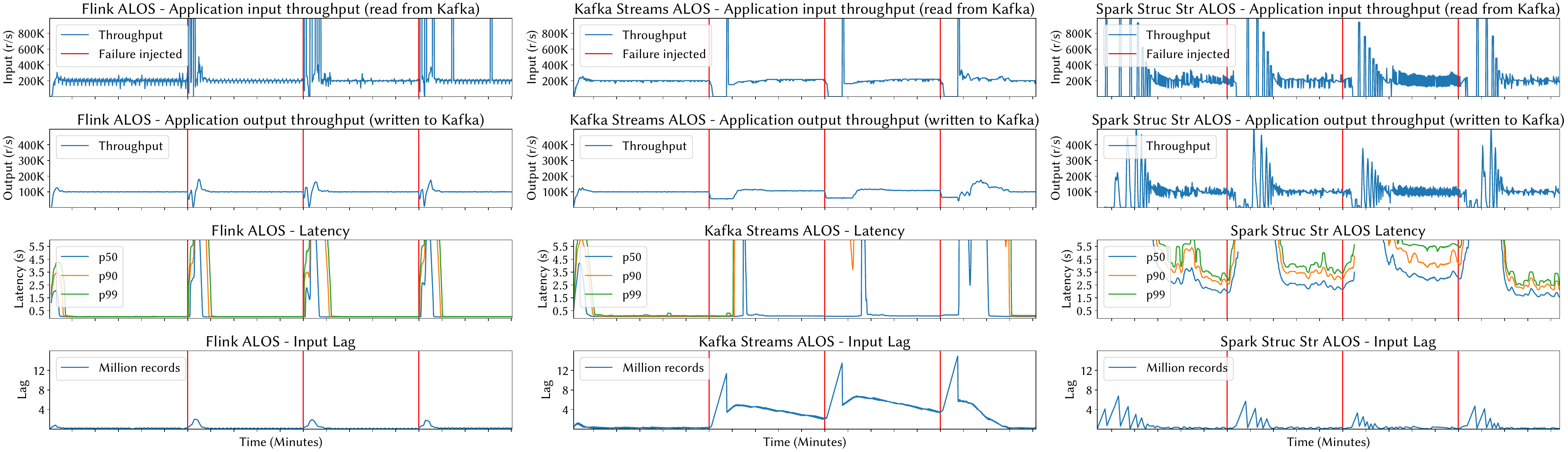}
    \caption{Fault recovery}
    \label{fig:characterization2podskill_performance}
  \end{subfigure}
    \begin{subfigure}[htbp]{1\textwidth} 
    \centering
    \includegraphics[width=\textwidth]{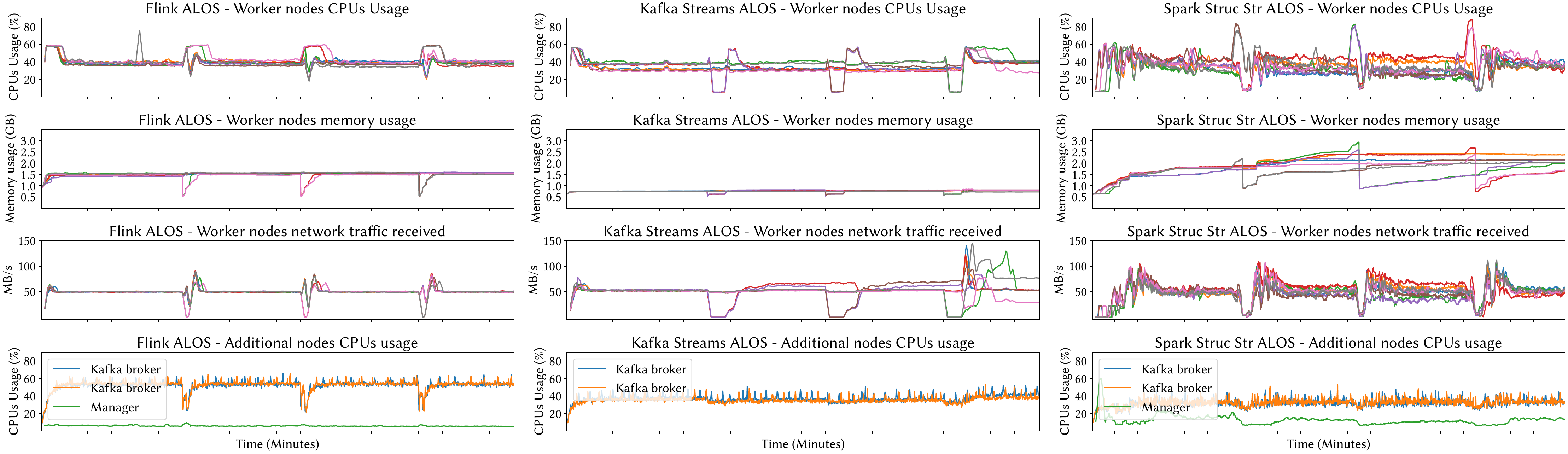}
    \caption{Resources utilization}
    \label{fig:characterization2podskill_resources}
  \end{subfigure}
  \caption{ALOS recovery characterization: Two random pods executing worker instances are killed, repeated 3 times.}
  \label{fig:characterization2podskill}
\end{figure*}

\textbf{Resource consumption observation 1}: \textit{Spark Structured Streaming has the most dynamic CPUs usage}. This is due to its modern execution model based on fine-grain tasks assigned to Spark's executors (worker instances). Tasks are created within the so-called \textit{stages} that execute the micro-batches. The number of tasks is defined based on the number of partitions (consumed from Kafka and shuffle partitions between the DAG's stages). Such a dynamic execution model can be more scalable and provide load balancing, but it comes at the cost of potential scheduling overheads due to assigning all the fine-grained tasks to executors. From the operating system view, Spark Structured Streaming execution is the one that allocates the largest memory space.\footnote{Considering that how Java Virtual Machine allocates memory and computes can vary across the frameworks, this memory observation should not be taken as evidence because the operating system's view of memory usage does not guarantee accuracy.} Considering that memory consumption can heavily impact performance, this motivates future analysis to comprehend the root causes and potential mitigations. One direction is to measure the garbage collection (GC) effectiveness and performance, where Spark's GC was fine-tuned in related work~\cite{vanDongen2021a}. Moreover, this behavior could be a potential reason for Spark's performance limits found in our previous work~\cite{ICPE2024}.

\textbf{Resource consumption observation 2}: From a resource consumption perspective, \textit{Kafka Streams shows unstable behavior after the failures are injected}. This is supported mostly by varied CPU usage and network traffic in the worker nodes, which suggests some imbalances aligned with \textbf{performance~observation~2}. Furthermore, Kafka Streams' executions allocate the least memory, which is a positive outcome from a resource efficiency view.  

\textbf{Resource consumption observation 3}: \textit{Flink shows consistent resource utilization under failures}. Although the failures cause fluctuations, the executions return to a normal stable state. It is worth noting that Flink's execution is the one causing more CPU usage in the large instances of Kafka Broker servers. This can be due to Flink's fine-grained event-based processing, which continuously generates outputs to the Kafka topic. We noted that Flink's performance may rely more than other frameworks on the behavior of the Kafka brokers. We intend to explore this insight further in a future analysis.

\Cref{fig:characterization2podskill} shows the scenario where two random pods are periodically killed. We expect this scenario of more workers failing to be more challenging as more internal state is lost and more records need to be reprocessed by being replayed from the input source. Compared to \cref{fig:characterization1podkill}, \cref{fig:characterization2podskill} shows similar performance, fault recovery, and resource consumption observations. However, it is noticeable in \cref{fig:characterization2podskill} that Kafka Streams' execution accumulates more lag after the failures and its event latency increases dramatically.

\begin{figure*}[htbp]
  \centering
  \begin{subfigure}{1\textwidth} 
    \centering
    \includegraphics[width=\textwidth]{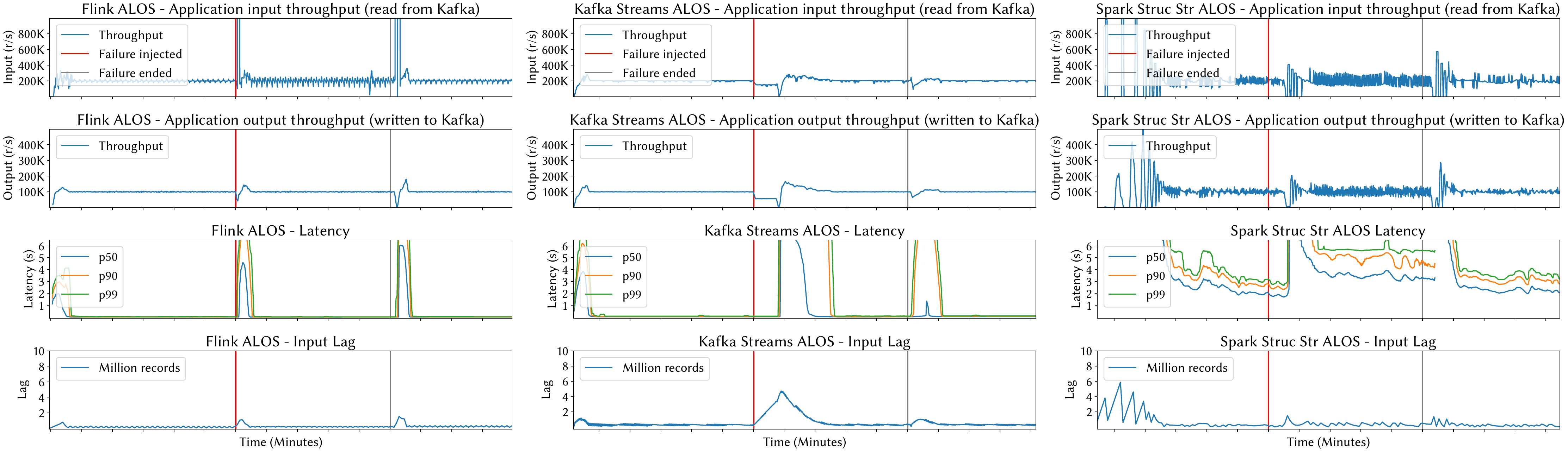}
    \caption{Fault recovery}
    \label{fig:characterization2podscrash_performance}
  \end{subfigure}
    \begin{subfigure}{1\textwidth} 
    \centering
    \includegraphics[width=\textwidth]{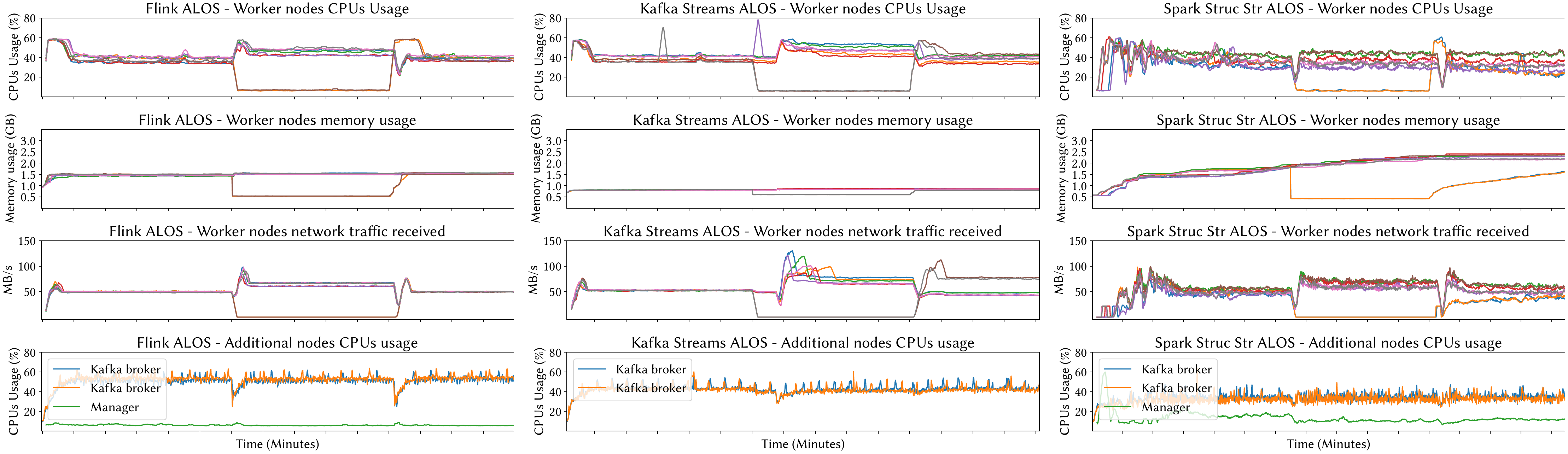}
    \caption{Resources utilization}
    \label{fig:characterization2podscrash_resources}
  \end{subfigure}
  \caption{ALOS fault recovery characterization: Two random pods executing worker instances crash for 5 minutes.}
  \label{fig:characterization2podscrash}
\end{figure*}

Kafka Streams' execution shows good performance after failures in some cases, e.g., after the third failure demonstrated in \cref{fig:characterization2podskill} where the rebalance enables the lag to be processed quickly and the latency reduced. \Cref{fig:characterization2podskill_resources} shows that a Kafka Streams' worker node (indicated in green) consumes more CPUs and has higher network traffic while the lag is being reduced. This indicates that a worker instance pod on this specific node was assigned to a lagged partition and processed the lagged records. On the other hand, there are many cases where the distribution of partitions is suboptimal. For example, \cref{fig:characterization2podskill} shows that after the first failure, some workers take too long to process the accumulated lag compromising the recovery. Therefore, \textbf{Fault~recovery~observation~2}  is that \textit{Kafka Streams fault recovery's impact on application performance varies across recurrent failures}.

We do not show here the results from four pods being killed because they are aligned with the insights from the failures of two worker instances~\cite{ReplicationPackage}. \Cref{fig:characterization2podscrash} shows the scenario of failure injected where random pods executing worker instances crash (see more in \cref{sec:scenarios}). When pods become unavailable, the local state is lost. The execution must continue processing with fewer worker instances, as new pods are not created when pods crash. 

Significant drops in system utilization are noticeable after the failure in the machines executing the two crashed pods~\cref{fig:characterization2podscrash_resources}. Moreover, in the three frameworks, the remaining active workers increased their resource utilization to cope with the failure. When the injected failure is halted (the gray line), the executions can return to the normal state with crashed worked instances being re-added to the worker pool. From the performance and QoS standpoint evinced in \cref{fig:characterization2podscrash_performance},  \textbf{Fault~recovery~observation~3} is that \textit{the executions of the three frameworks recovered from a critical failure with fewer resource available, they also returned to a comparable state as the one before the failure when the failure is halted.} 

One can note in \cref{fig:characterization2podscrash_performance} that Flink's execution is the one that recovers faster. Kafka Streams is the one accumulating more lag and its execution took almost 3 minutes to reduce the latency. After the failure is halted, Kafka Streams ends up with some noticeable imbalances in network traffic and CPU usage (\cref{fig:characterization2podscrash_resources}). 

\Cref{fig:characterization2podscrash_performance} shows Spark Structured Streaming with fluctuations after the failure, which is due to the reprocessing of the failed executors' tasks and its dynamic micro-batching. The micro-batching interval was increased to a higher value of around 3 seconds, attempting to cope with the same data volume and fewer processing resources. However, this caused a slight increase in Spark Structured Streaming's latency and fluctuations in the output throughput.

\subsection{EOS Recovery Characterization}\label{sec:characterization_eos}

\begin{figure*}[htbp]
    \includegraphics[width=.8\textwidth]{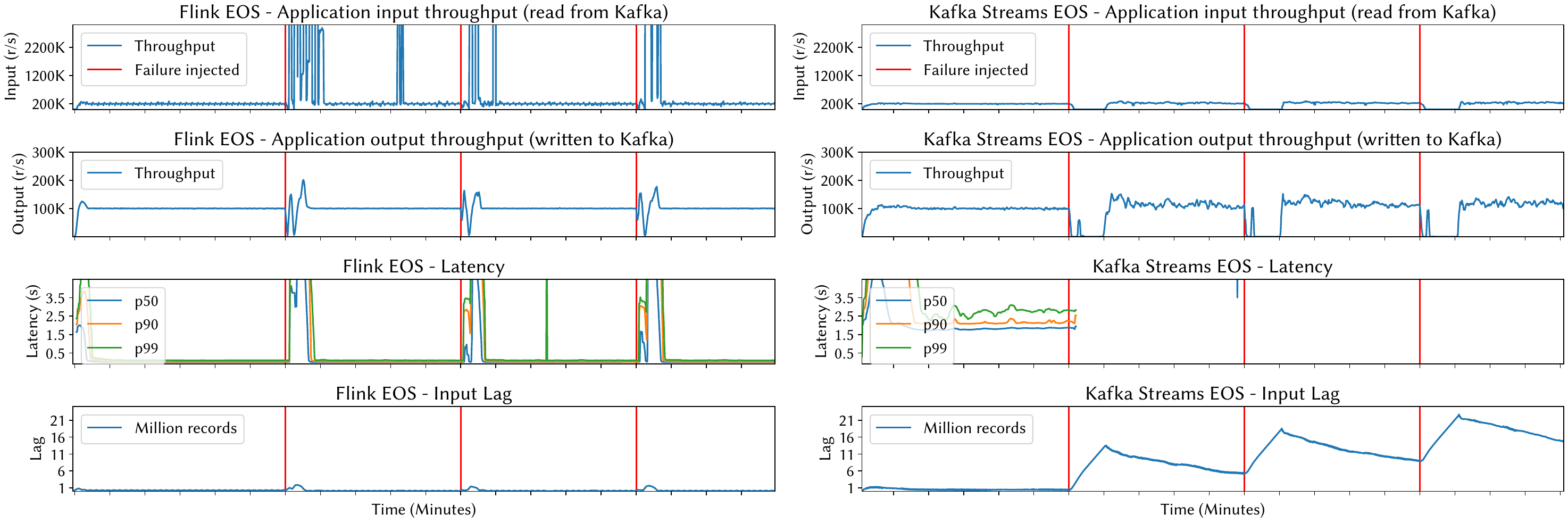}
    \label{characterization_eos_1pod-kill_performance}  
  \caption{EOS recovery characterization: Random pods executing worker instances are killed, repeated 3 times every 5 minutes.}
  \label{fig:characterization_eos_1podkill}
\end{figure*}

Supporting exactly-once semantics (EOS) requires additional deployment considerations. One more Kafka broker was needed for resource stability and the fault tolerance guarantee had to be enabled in the frameworks. Moreover, the isolation level was configured to read only committed data to prevent the reading of duplicated messages in downstream operators, consistent with practices in related literature~\cite{vanDongen2021a,Wang2021}.
Considering that Spark Structured Streaming can not natively guarantee exactly-once when writing the output to a Kafka topic, we do not show results from Spark Structured Streaming on exactly-once experiments, which is consistent with the related literature~\cite{vanDongen2021a}. Flink provides EOS output (also known as output commit problem) using the two-phase commit protocol and Kafka Streams with custom sinks for rollback-recovery~\cite{carbone2017, Fragkoulis2023}.

\begin{figure*}[htbp]
  \centering
  \begin{subfigure}[htbp]{.3307\textwidth} 
    \centering
    \includegraphics[width=1\textwidth]{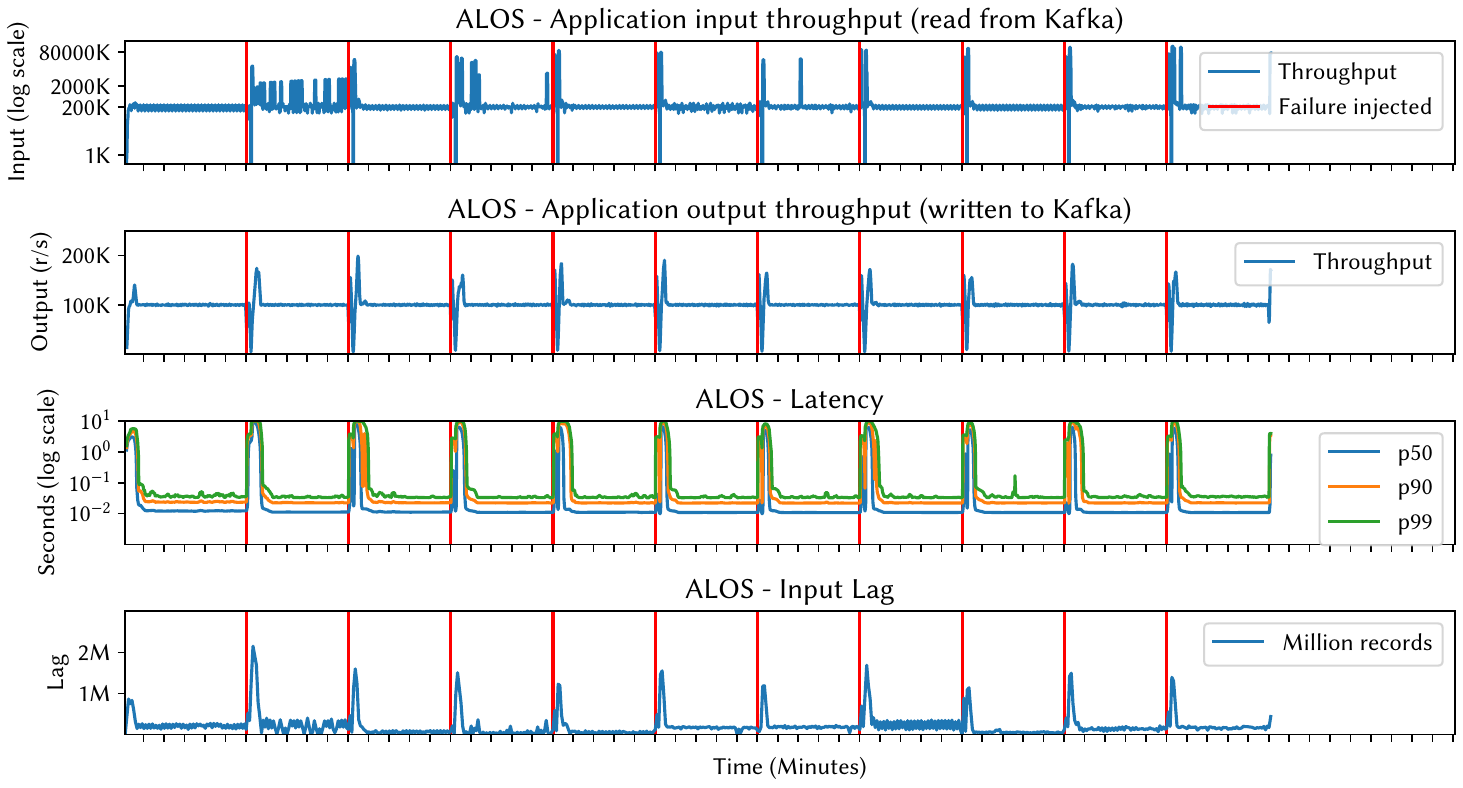}
    \caption{Flink}
    \label{fig:flink1killed}
  \end{subfigure}
  \begin{subfigure}[htbp]{.3307\textwidth} 
    \centering
    \includegraphics[width=1\textwidth]{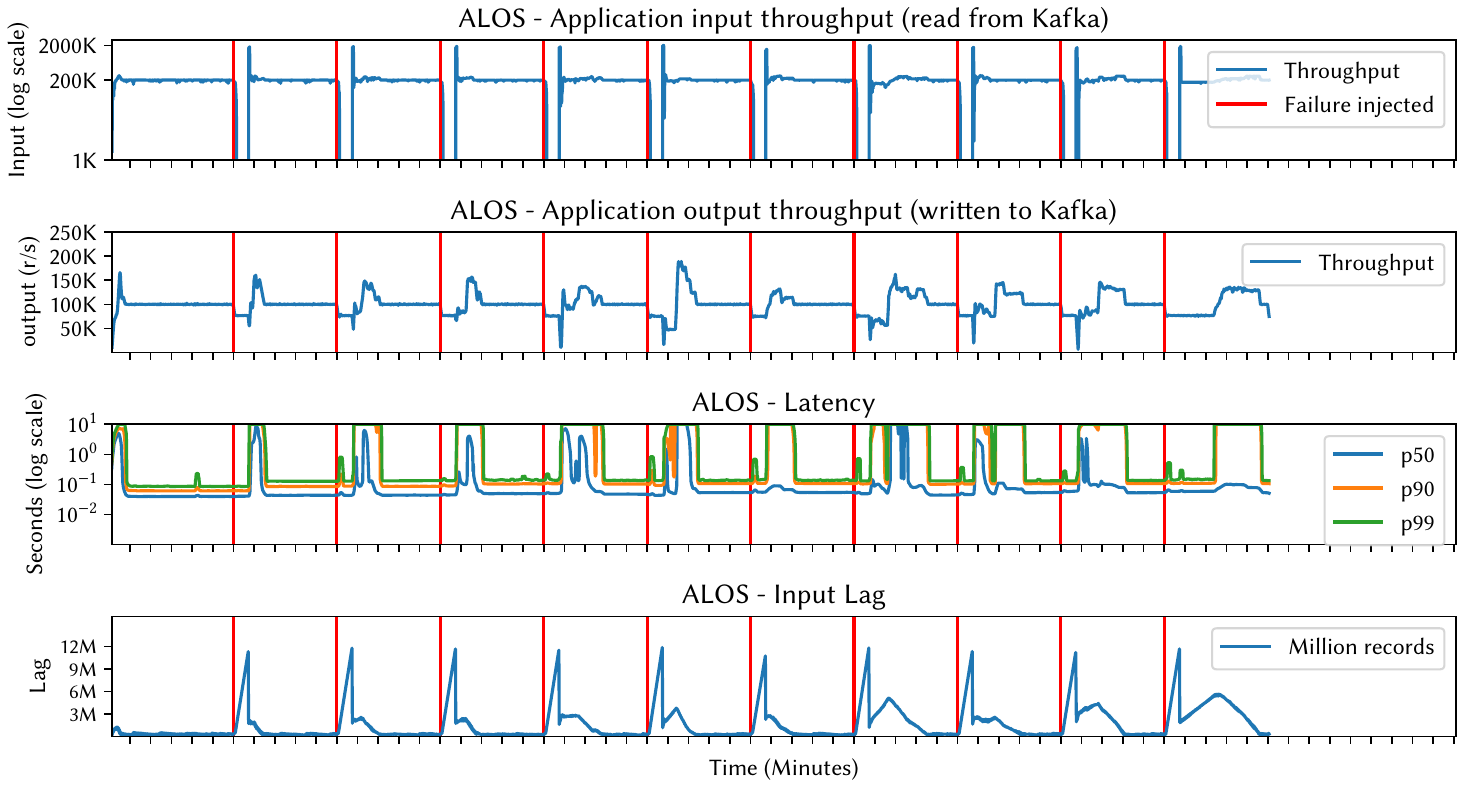}
    \caption{Kafka Streams}
    \label{fig:kstreams1killed}
  \end{subfigure}
  \begin{subfigure}[htbp]{.3307\textwidth} 
    \centering
    \includegraphics[width=1\textwidth]{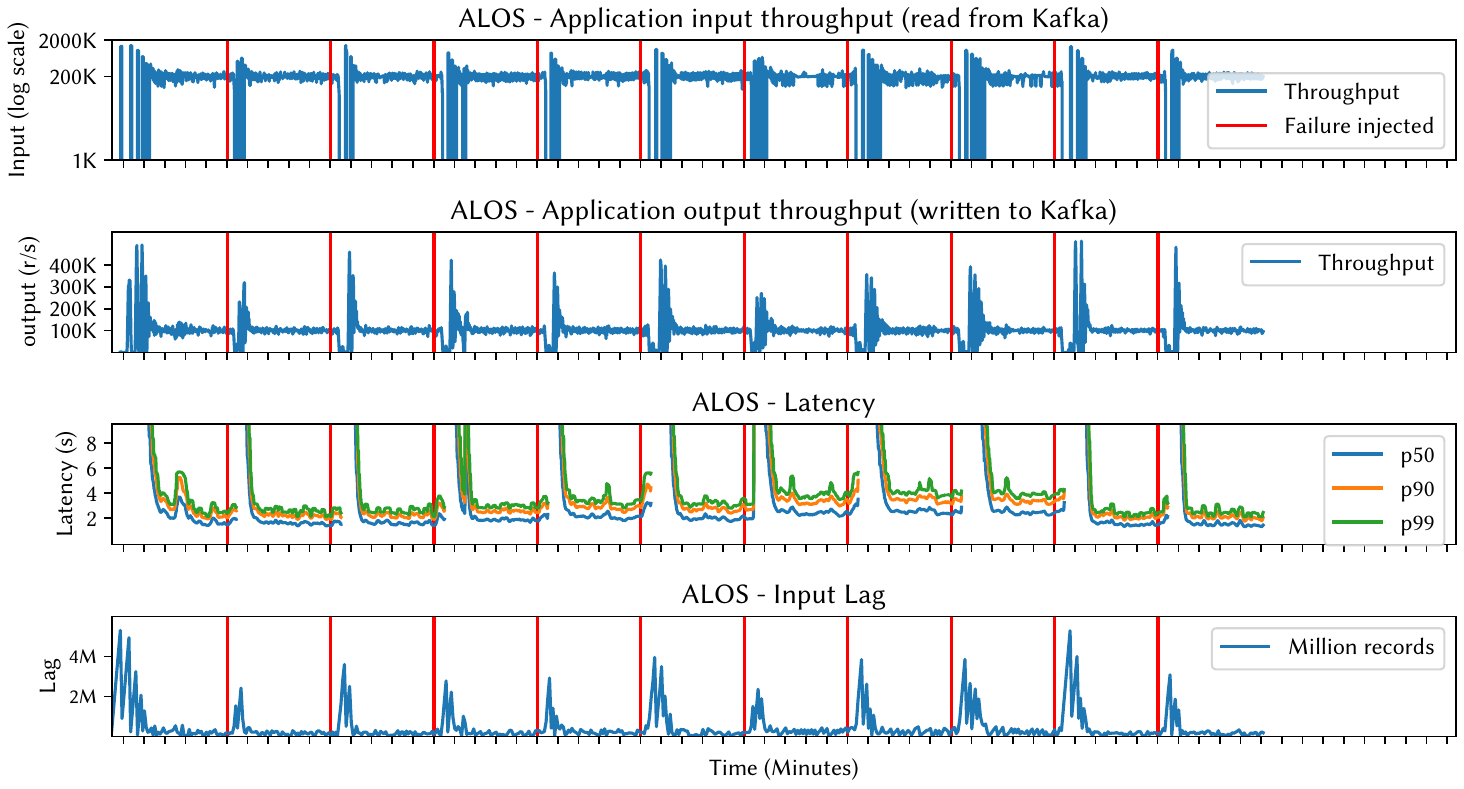}
    \caption{Spark Structured Streaming}
    \label{fig:spark1killed}
  \end{subfigure}
  \caption{ALOS fault recovery: Random pods executing worker instances are killed, repeated 10 times every 5 minutes.}
  \label{fig:stability_1worker_kill}
\end{figure*}

\begin{figure*}[htbp]
  \centering
  \begin{subfigure}[htbp]{.3307\textwidth} 
    \centering
    \includegraphics[width=1\textwidth]{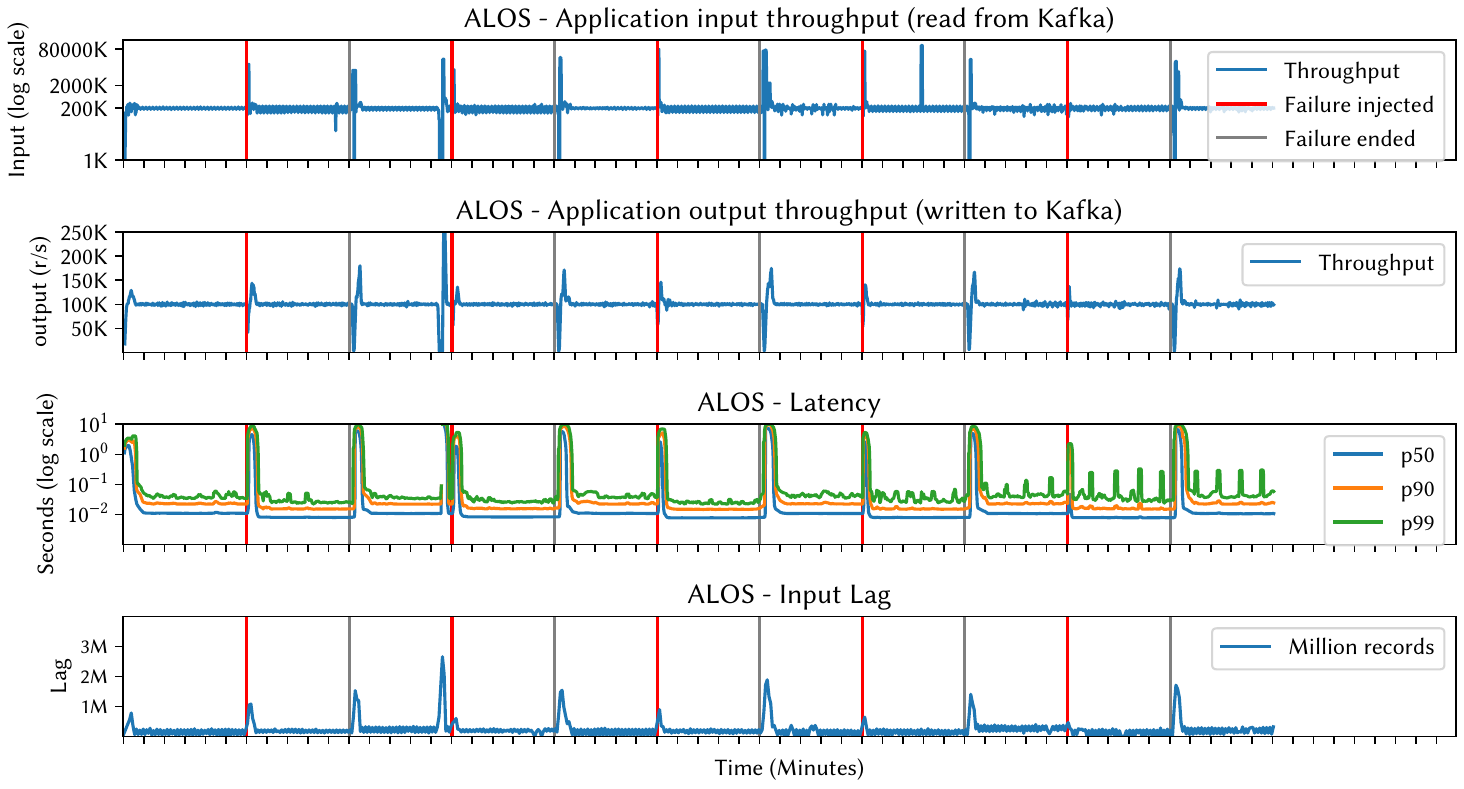}
    \caption{Flink}
    \label{fig:flink2crash}
  \end{subfigure}
  \begin{subfigure}[htbp]{.3307\textwidth} 
    \centering
    \includegraphics[width=1\textwidth]{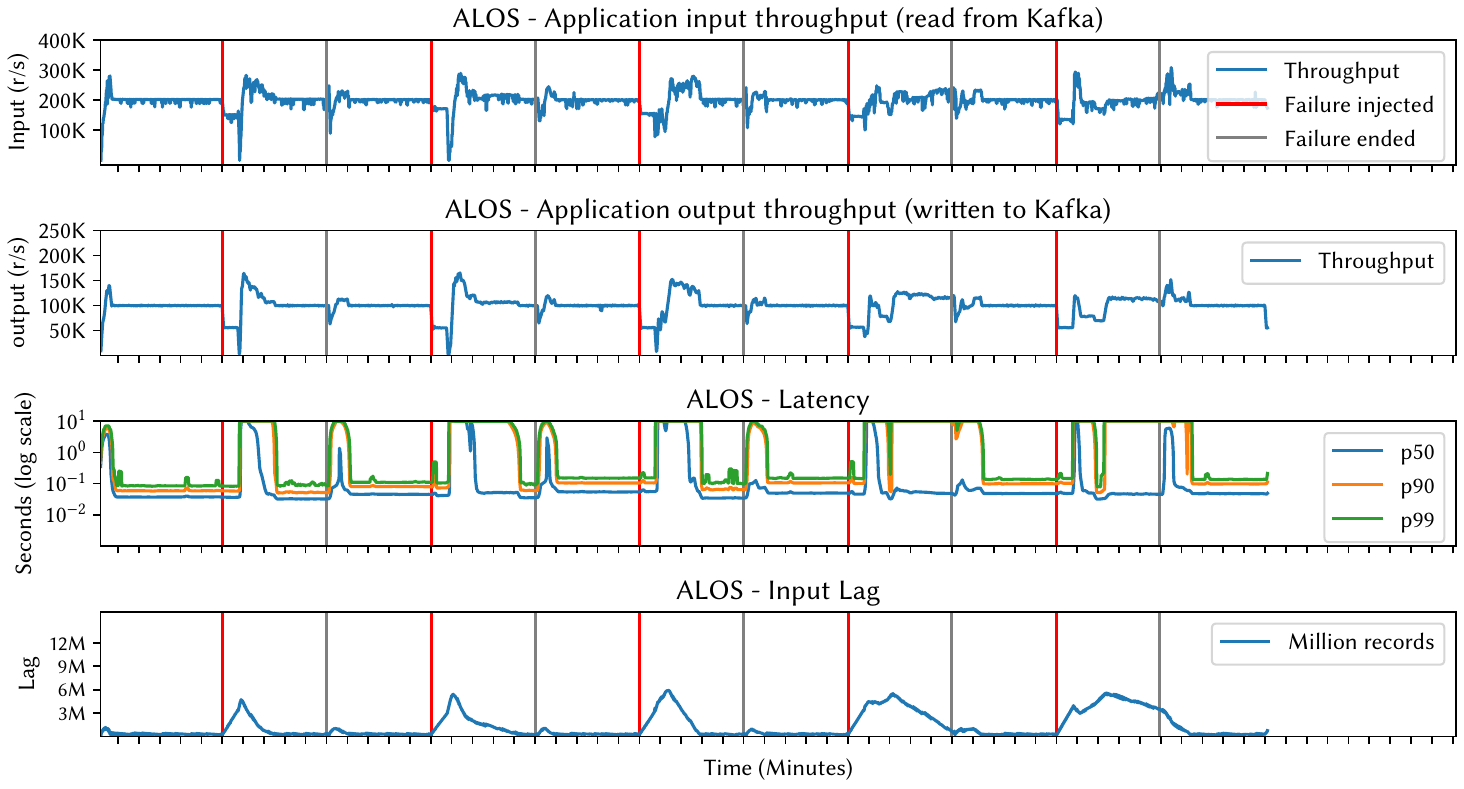}
    \caption{Kafka Streams}
    \label{fig:kstreams2crash}
  \end{subfigure}
  \begin{subfigure}[htbp]{.3307\textwidth} 
    \centering
    \includegraphics[width=1\textwidth]{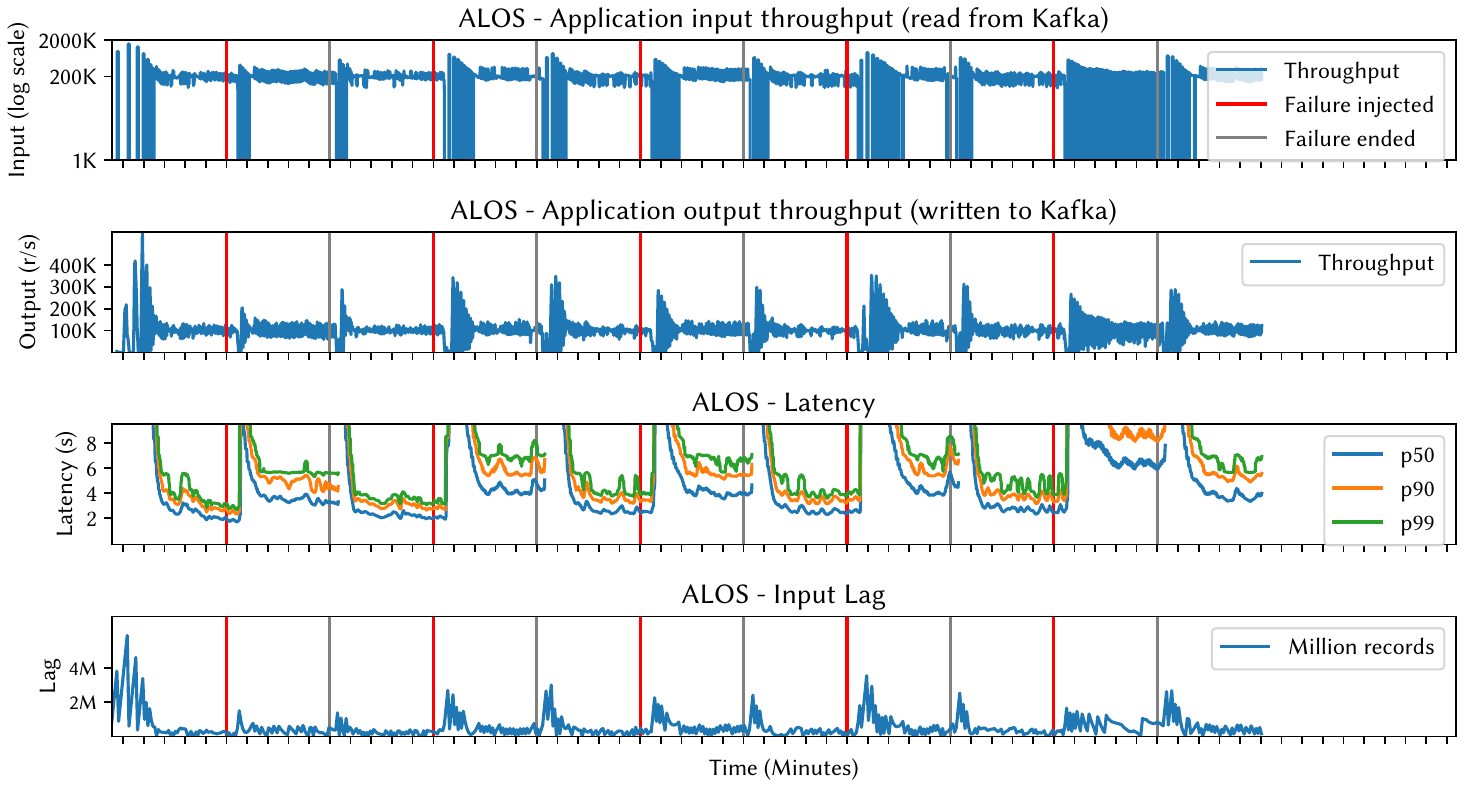}
    \caption{Spark Structured Streaming}
    \label{fig:spark2crash}
  \end{subfigure}
  \caption{ALOS recovery: Two random pods executing worker instances crash for 5 minutes, repeated 5 times every 10 minutes.}
  \label{fig:stability_workers_crash}
\end{figure*}

We show only the results with one pod kill that already provides the insights necessary to highlight Flink's advantages and Kafka Streams' shortcomings when EOS is needed.
In~\cref{fig:characterization_eos_1podkill} it becomes evident that Flink provides better performance and fault recovery than Kafka Streams.~\footnote{System monitoring results are not included in this section for the sake of conciseness, as there are no new insights from a resources perspective.}
One can note that Flink achieved a much lower latency in \cref{fig:characterization_eos_1podkill}. Kafka Streams stabilized in a latency above 2 seconds and the latency does not normalize after the failure due to the significant lag. After failures, Kafka Streams' latency is invisible in the y-range as it exceeds the 10-second bucket limit~\cite{ICPE2024}. The higher latency is due to the records buffering in Kafka Streams' repartitioning topic.

Both frameworks' executions are similar in the throughput until the first failure occurs. After the failure, Flink recovered faster in less than a minute. Kafka Streams took longer to reach a higher output throughput that remained unstable due to the accumulated lag and slow processing, which aligns with \textbf{Performance~observation~2} and \textbf{Fault~recovery~observation~2}. Given that the scenario requiring exactly-once semantics is more critical, Kafka Streams' suboptimal rebalance strategy has a greater impact. Thus, \textbf{Fault~recovery~observation~4} is that \textit{Flink provides better performance and fault recovery for use cases that demand exactly-once guarantees.}

\section{Fault Recovery Stability} \label{sec:stability}

This section presents the results of fault recovery stability under recurrent failures injected. 
For the sake of visual clarity, note the different ranges and scales on the y-axis. For conciseness, we limit the discussion to the results with ALOS guarantees as those are the most complete and insightful.


\Cref{fig:stability_1worker_kill} complements the results from \cref{fig:characterization1podkill} with a focus on the fault recovery stability. \cref{fig:flink1killed} shows that despite some fluctuations in the input throughput of Flink's execution, it achieved a stable behavior under recurrent failures.
\Cref{fig:kstreams1killed} shows Kafka Streams execution in a scenario where one random worker instance is recurrently killed. Focusing on throughput and lag, it becomes evident that Kafka Streams failure recovery follows a pattern with different phases after the failures and it lasts for several minutes.

 Although Kafka Streams' failure recovery pattern is consistent across the different failures, it varies significantly for how long the patterns and their phases occur, and their impact on performance (e.g., lag peaks and throughput drops). Moreover, the fluctuations in lag and throughput have a detrimental impact on the latencies, evincing limitations in Kafka Streams' fault recovery. On the one hand, in the recovery of the sixth failure shown in \ref{fig:kstreams1killed}, the lag accumulated as well as the throughput are lower, and the latency increase is not disruptive with the p50 being kept within the boundaries. On the other hand, following the seventh failure, the performance degradation becomes critical. Consequently, the execution takes around 4 minutes to return to a state comparable to the one before the failure. 

In terms of reasons for the observed Kafka Streams' fault recovery instabilities, the current understanding is that it is mostly due to which worker instance the failure is injected (e.g., if the consumer group leader is affected or not) and how the subsequent rebalances are enacted. On the one hand, this outcome is aligned with \textbf{Fault~recovery~observation~2}. On the other hand, we believe that this potentially problematic behavior of Kafka Streams should be further characterized and improved.

\Cref{fig:spark1killed} shows Spark Structured Streaming's behavior under recurrent failures. On the one hand, there are some minor noticeable fluctuations in the lag accumulated, latencies after failures (lowest latencies in the last failures injected), and throughput recovery times. These minor fluctuations are justified by the dynamic micro-batching and the fine-grain tasks mapping to the worker executor instances. On the other hand, the impact on performance and QoS is similar across the recurrent failures, which indicates that Spark Structured Streaming can also provide stable fault recovery.

\textbf{Fault~recovery~observation~5} is that \textit{Flink and Spark Structured Streaming can provide stable fault recovery. Kafka Streams, on the contrary, has unstable behavior with volatile recovery time.}

\Cref{fig:stability_workers_crash} complements the results from \cref{fig:characterization2podscrash}. If we compare the recovery from the first failure shown in \cref{fig:characterization2podscrash} to the five failures covered in \cref{fig:stability_workers_crash}, one can note that Flink and Spark Structured streaming recoveries are similar in all failures. Kafka Streams, on the contrary, result in failures having again different impacts on its performance, mostly in terms of latency, lag accumulation, and lag reduction. We can then conclude that \textbf{Fault~recovery~observation~2} also applies to worker instances crashes.

\section{Fault Recovery Time}\label{sec:recovery-time}

Estimating fault recovery time is another relevant metric that complements the results from \cref{sec:characterization} and \cref{sec:stability}. Considering that there is a lack of approaches and tools to automate the measurements of fault recovery times across multiple failures, we designed a simple detector based on moving average and standard deviation to automatically detect the failure as well as the time to recovery. The detector extracts the expected average and standard deviation for the throughput based on the normal processing behavior. We define the normal processing period as the interval between the execution warm-up and the injection of the first failure, which can vary depending on the time required for warm-up and the timing of the first failure. In our analysis, the normal processing period is between the second and fifth minute after the start of execution across all frameworks. 

In short, the detector marks a point of failure injection when the moving average of a window of 10 data points and its standard deviation are two times above the one from the normal processing period. The standard deviation threshold and window size are parametric values defined for all frameworks. Afterward, another point of failure was strictly marked every 5 minutes, which is the interval between failures used in our experiments. 

\begin{figure}
    \centering
    \includegraphics[width=1\linewidth]{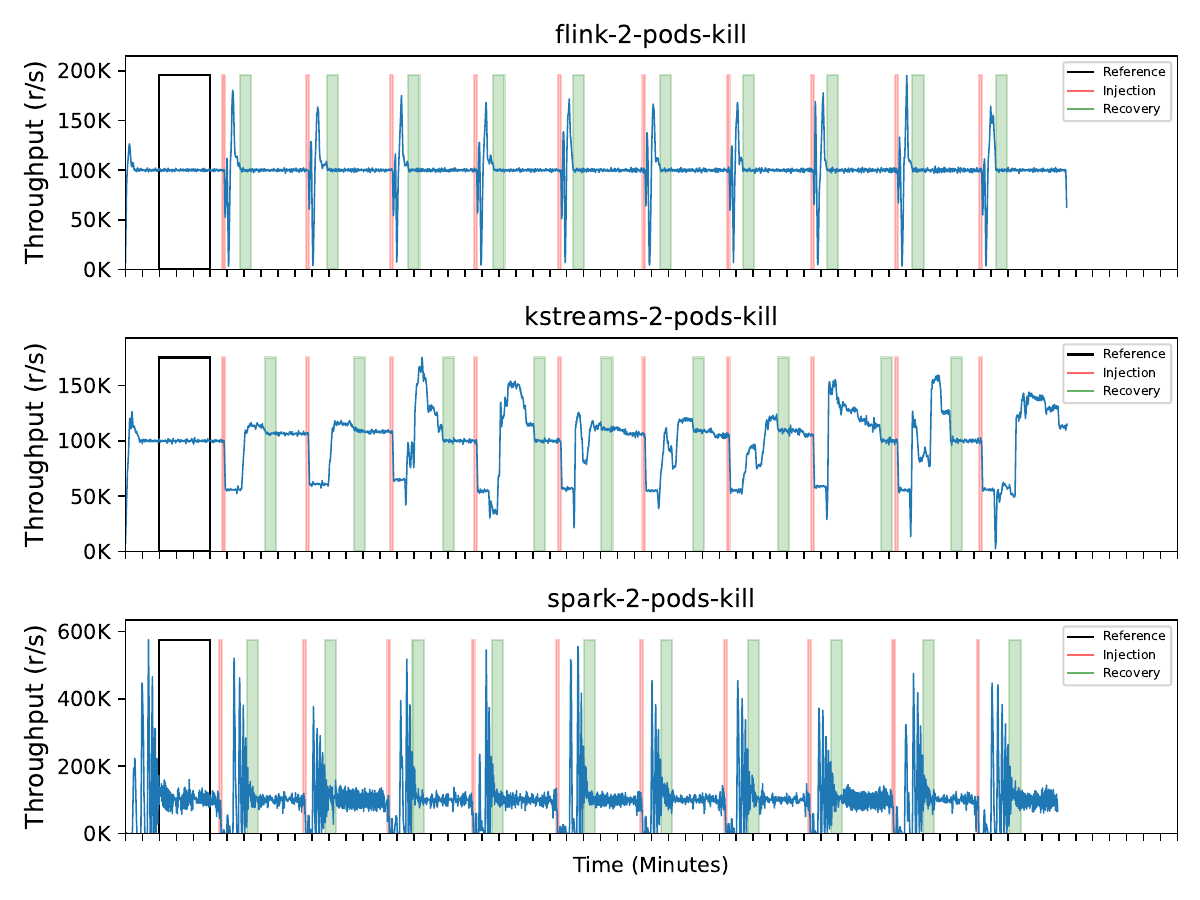}
    \caption{Characterization of our approach to automate measurements of fault recovery times in the 2-pods kill ALOS scenario. The stable behavior, used as a reference for training, is represented by a black rectangle. Failures are highlighted in red, while recovery detection is highlighted in green.}
    \label{fig:recovery-approach-characterization}
\end{figure}

The recovery time was detected when the values returned within a threshold of 15\% for a time frame of at least 40 seconds. We expect throughput to stabilize after the recovery time, which should be only detected when the execution returns to a normal state. It is important to note that our approach for detection failures and their recovery was configured to automate our measurements. Considering that it is parametric and customizable, we expect it to be also applicable to wider scenarios.\footnote{The tooling and all the parameters used are available as supplementary material~\cite{ReplicationPackage}.} Moreover, we believe that there are no ground truths about these parameter values used. In fact, we focus more on having a formal parametric method to automate the measurements that is less arbitrary and avoids biased analysis. The values detected can be an estimation of the recovery time that we expect to be ascertained with visual inspections in the metrics traces. \Cref{fig:recovery-approach-characterization} exemplifies the trace of automatically detected failures (of 2-pods kill) and their recoveries.

\begin{figure}[htbp]
  \centering
    \includegraphics[width=1\linewidth]{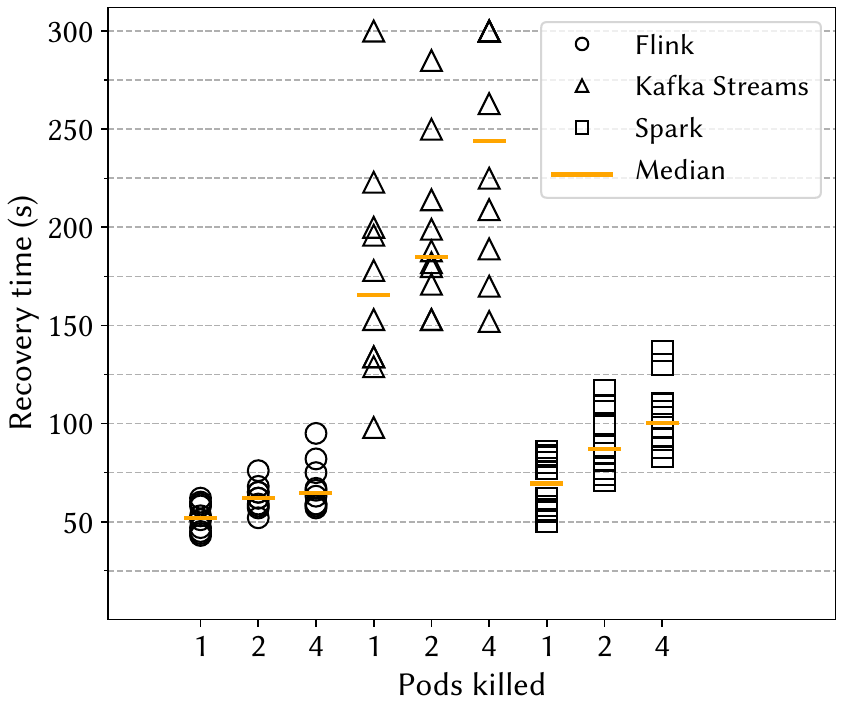}
  \caption{Output throughput recovery times by framework.}
  \label{fig:recovery-times}
\end{figure}

\Cref{fig:recovery-times} summarizes the results of the fault recovery times related to the application output throughput. Noteworthy, Flink achieves the fastest overall recovery time for both metrics. Kafka Streams' recovery times are longer and with higher variability, which aligns with the discussion provided in~\cref{sec:characterization,sec:stability}. 

In the Kafka Streams scenario of 4 pods killed, a complete recovery was not achieved in the 5-minute window in four out of ten injected failures. The same event happened with one Kafka Streams failure in the scenario of 1 pod being killed. In such cases, the fault recovery time is defined as the time interval between failures. Defining the fault recovery time for these cases as the failure injection interval can impact the estimation because a complete recovery could take longer.\footnote{The recovery accuracy can be improved with longer intervals between failures, where our follow-up paper provides preliminary results from such a scenario~\cite{vogel2024}.} Despite such a scenario, our method captured Kafka Streams' fault recovery as the slowest due to its limited fault recovery.

Generally, \textbf{Fault~recovery~observation~6} is that \textit{Flink and Spark Structured Streaming provide faster fault recoveries, and Flink demonstrates the highest level of stability}. Considering that fault recovery time is also relevant in terms of the latency metric, we are currently measuring it, and preliminary results are already available in our follow-up paper~\cite{vogel2024}. 

\section{Results Discussion}

Our comprehensive experimental method demonstrates that stream processing frameworks significantly impact application performance and fault recovery. Given that fault recovery significantly affects the quality of service, it is a crucial criterion when selecting a stream processing framework for production deployment. ~\Cref{sec:observations} shows our key insights from the experiments and \cref{sec:findings} discusses the main findings.

\subsection{Observations Summary and Takeaways}\label{sec:observations}

The following are key insights from our experiments and the observations discussed in~\cref{sec:characterization,sec:stability,sec:recovery-time}:

\begin{itemize}

    \item[--] Flink, Kafka Streams, and Spark Structured Streaming executions are resilient to recover from failures of killed and crashed worker instances, even from failures affecting many worker instances. Yet, the impact of their recoveries on the applications' performance varies in terms of the application's performance metrics, recovery time, and how stable the recoveries are across recurrent failures.
    
    \item[--] In use cases where at-least-once guarantees are suitable, Flink and Spark Structured Streaming provide a consistently fast and stable recovery. Spark Structured Streaming can be a suitable alternative when low latency is not required. On the other hand, Kafka Streams takes longer to recover from failures and has volatile behavior.
    
    \item[--] Flink is the most suitable framework for use cases requiring exactly-once semantics (EOS) guarantees, providing fast recovery and stable performance. Conversely, Kafka Streams' executions have lower performance when EOS is enabled, and Spark Structured Streaming has limited native support for guaranteeing EOS in executing applications.

\end{itemize}

\subsection{Findings discussion}\label{sec:findings}

Although many of the insights we discussed regarding how stream processing frameworks recover from failures align with related literature~\cite{vanDongen2021a,Wang2021,Wang2022}, our experiments reveal new trends and insights that warrant consideration. To the best of our knowledge, the following are key novel findings:

\begin{itemize}

  \item[$\star$] \textbf{Finding}: Kafka Streams provides limited fault recovery.
  \item \textbf{Rationale}: The current Kafka Streams' partition assignment strategy has not been thoroughly evaluated in related approaches~\cite{vanDongen2021a,Wang2021}. Kafka Streams was only tested under failures in~\cite{vanDongen2021a} with a single failure injected. Our experimental method reached this finding due to the recurrent injection of different failures and recovery time measurements.
\end{itemize}

\begin{itemize}
  \item[$\star$] \textbf{Finding}: Flink currently provides fast and stable recovery.
  \item \textbf{Rationale}: Measuring the stability across recurrent failures is a novelty of our experimental method, which also covers more recent enhancements implemented on Flink. Relevant enhancements in Flink include the reactive mode, as well as optimizations in checkpointing and load management~\cite{Fragkoulis2023}.
\end{itemize}

\section{Conclusions}

Ensuring fault tolerance in data-intensive, event-driven applications is crucial for successful industry deployments. In this paper, we contribute with a more comprehensive evaluation method and fault recovery analysis. Such an evaluation was enabled by \emph{ShuffleBench's}~\cite{ICPE2024} flexible deployment, parameters, and extension to fault recovery metrics. 

\subsection{Results Implications}

One of the main implications of our findings is that Kafka Streams, despite its growing interest, needs improvements in terms of fault recovery. This implication is relevant because our findings indicate that in contrast to what was reported in the past~\cite{vanDongen2021a}, Kafka Streams has unstable recovery that can compromise QoS and service level agreements. The finding that Kafka Streams faces performance instabilities after failures due to its current rebalancing strategy has relevant implications. We intend to share our findings with Kafka Streams' community and contribute with future enhancements. 

The positive recovery results seen in Flink imply that it is currently the optimal solution for what mostly motivates fault tolerance: resilience and continuous availability of services. 
Another relevant implication from the fault recovery stability measurements is that the common assumption that frameworks' recovery performance is comparable before and after recovery~\cite{vanDongen2021a} does not hold. In fact, our results indicate significant contrasts in recoveries across failures (see \Cref{sec:recovery-time}), which can impact the applications' performance, QoS, and user experience.

The insights from our experimental results can also impact on advanced optimizations necessary in modern stream processing, such as run-time self-adaptations for auto-scaling. This connection arises because implementing run-time adaptations in distributed executions inherently involves fault recovery procedures. Considering that our experimental method is comprehensive and includes insights into how chaos engineering can be applied to long-running computing systems, it is expected to be generic enough to be useful for evaluating other domains.

\subsection{Limitations and Threats to Validity}

We expect all threats to the internal validity discussed in our previous works to also apply to our results~\cite{ICPE2024}. 
In addition to known threats to external validity, the fault measurements have additional configurations and execution threats. The failure injection interval was defined according to insights from real-world industry deployments, where there can be hundreds or thousands of worker instances, thus increasing failure likelihood. However, we still need more comprehensive statistical studies to provide insights into failure rates in large-scale cloud deployments. From our standpoint, we believe that new comprehensive studies are needed to characterize the failure rates in modern public hyperscalers, similar to what we have seen in high-performance computing centers~\cite{schroeder2009}.

\subsection{Future Work}
We expect that future analyses to extend ours are still needed.
With the significant impact of fault recovery on applications' performance and QoS, we argue that modern performance benchmarking should be extended to support more realistic scenarios. Performance is still a major goal, but it should be continuously improved. This would be a starting point to better representing real-world scalable computing deployments in clouds.

A relevant lesson learned from the experiments is that 5 minutes between two injected failures is not enough to measure the recovery time in all scenarios. Therefore, future work is to experiment with failures injected over longer intervals. We also believe the following are necessary complementary analyses:
\begin{itemize}
    \item[--] Cover other types of failures, such as in the manager node.
    \item[--] Extend our method for workloads with larger state sizes.
    \item[--] Evaluate the impact of the frameworks on the Kafka brokers' performance, resource utilization, and costs.
    \item[--] Extended our analysis of fault recovery for more scenarios where EOS guarantees are needed.
    \item[--] Further analysis of the frameworks' output correctness under failures with different fault tolerance semantic guarantees. 
    \item[--] Extend the measurements to cover execution costs with the stream processing frameworks. 
    
\end{itemize}

\begin{acks}
We would like to thank Johannes Kepler University Linz and Dynatrace for co-funding this research. We thank Bernhard Kepplinger for providing insights into the method for automating the measurement of failure recovery times.
\end{acks}

\bibliographystyle{ACM-Reference-Format}
\balance
\bibliography{submission}

\end{document}